\documentclass[doublecol]{epl2}

\title{Entanglement temperature in molecular magnets composed of $S-$spin dimers}

\author{D. O. Soares-Pinto\inst{1,3} \and A. M. Souza\inst{2} \and R. S. Sarthour\inst{3} \and I. S.
Oliveira\inst{3} \and M. S. Reis\inst{4} \and P.
Brand\~{a}o\inst{4,5} \and J. Rocha\inst{4,5} \and A. M. dos
Santos\inst{6}} \shortauthor{D. O. Soares-Pinto \etal}

\institute{
  \inst{1} Instituto de F\'{i}sica de S\~{a}o Carlos,
Universidade de S\~{a}o Paulo, P.O. Box 369, 13560-970, S\~{a}o
Carlos - SP, Brazil.\\
  \inst{2} Institute for Quantum Computing and Department of Physics and Astronomy, University of Waterloo,
Waterloo, Ontario, N2L 3G1, Canada.\\
  \inst{3} Centro Brasileiro de Pesquisas F\'{\i}sicas, Rua Dr.
Xavier Sigaud 150, 22290-180, Rio de Janeiro, Brazil.\\
  \inst{4} CICECO, Universidade de Aveiro, 3810-193, Aveiro,
Portugal.\\
  \inst{5} Departamento de Qu\'{i}mica, Universidade de Aveiro, 3810-193 Aveiro,
Portugal.\\
  \inst{6} NSSD, ORNL, Oak Ridge TN 37831-6475, USA.
}


\abstract{In the present work, we investigate the quantum thermal
entanglement in molecular magnets composed of dimers of spin $S$,
using an Entanglement Witness built from measurements of magnetic
susceptibility. An entanglement temperature, $T_{e}$, is then
obtained for some values of spin $S$. From this, it is shown that
$T_{e}$ is proportional to the intradimer exchange
interaction $J$ and that entanglement appears only
for antiferromagnetic coupling. The results are compared to
experiments carried on three isostructural materials:
KNaMSi$_{4}$O$_{10}$ (M$=$Mn, Fe or Cu).}

\begin{document}

\maketitle

\section{Introduction}

For about a decade it has been realized that quantum
entanglement is a valuable resource for quantum information
processing, since it allows forms of communication that are
classically impossible \cite{livro_nielsen, 2009_RMP_81_865}. 
However, until recently it was believed that the
phenomenon could not exist beyond atomic scale, due to the
interaction between the system and the
environment. Such interaction would lead to decoherence of
the quantum state, destroying  entanglement.

However, some theoretical works raised the possibility that solid
state systems could also exhibit quantum entanglement at finite
temperatures \cite{2001_PRL_87_017901,2006_NJP_8_140}. This
``thermal entanglement'' might be experimentally detected with the
help of some  observables, or ``witnesses'', that are related to
thermodynamical quantities, which could be directly measured
\cite{2006_PRB_73_134404, 2006_PRA_73_012110, 2007_PRB_75_054422,
2008_PRB_77_104402, 2009_PRB_79_054408, 2008_EPL_84_60003}. An
Entanglement Witnesses (EW), by definition, has a negative
expectation value for certain types of entangled states
\cite{1996_PLA_223_1, 2002_JMathPhys_43_4237, 2005_PhysRep_415_207,
2007_QIC_7_1}. The demonstration that quantum entanglement can
influence the behavior of thermodynamical properties of solids, such
as magnetic susceptibility \cite{2008_PRB_77_104402,
2009_PRB_79_054408, 2003_Nature_425_48, 2005_NJP_7_258,
2006_NJP_8_95, 2008_EPL_84_60003, 2009_PRA_79_052337}, shows that
quantum effects can be related to important macroscopic quantities.
These features have established that the study of entanglement in
solid state systems \cite{2008_RMP_80_517} based on the observation
of such EWs are helpful tools to quantum information and quantum
computation, since many proposals of quantum chips are solid state
based \cite{2001_Nature_410_789, 2002_Nature_416_406,
2005_PRL_94_207208, 2005_PRL_94_190501, 2007_JPD_40_2999,
2007_NatureNano_2_312}.

The class of materials known as molecular magnets
\cite{2008_NatureMat_7_179} are among those which can
exhibit thermal entanglement. In this class of materials, the
intermolecular magnetic interactions are extremely weak compared to
those within individual molecules. Thus a bulk sample, comprised by
a set of non-interacting molecular clusters, is completely described
in terms of independent clusters. The small number of coupled spins
is very convenient from the point of view of the theoretical
description, which can be made through analytical functions on a low
dimensional Hilbert space. From a physical point of view, a
molecular magnet can combine classical properties found in a
macroscopic magnet \cite{2008_NatureMat_7_179} and quantum
properties, such as quantum interference
\cite{2008_NaturePhys_4_277} and entanglement
\cite{2006_PRB_73_134404, 2006_PRA_73_012110, 2007_PRB_75_054422,
2008_PRB_77_104402, 2009_PRB_79_054408}. Recently, molecular magnets
have been pointed out as good systems to be used in high-density
information memories and also, due to their long coherence times
\cite{2007_PRL_98_057201}, in quantum computing devices
\cite{2001_Nature_410_789, 2002_Nature_416_406, 2005_PRL_94_207208,
2005_PRL_94_190501, 2007_JPD_40_2999, 2007_NatureNano_2_312}.

In this work, we investigate the quantum entanglement of molecular
magnetic materials with $S-$spin dimers, that can be theoretically
modeled by a Heisenberg-Dirac-Van Vleck (HDVV) Hamiltonian. The
magnetic susceptibility of these systems can be directly related
to an EW. Based on this model, we have obtained, for
different spin values, the temperature below which the quantum
entanglement exists, or the temperature of entanglement ($T_e$).
The results are experimentally verified using the
magnetic susceptibility data, from a previous work 
(see Ref.\cite{2009_JSolStatChem_182_253} for details), obtained for
three isostructural materials: the first composed of antiferromagnetic
Cu$-$dimers ($S=1/2$), the second composed of ferromagnetic
Fe$-$dimers ($S=2$), and the third composed of antiferromagnetic
Mn$-$dimers ($S=5/2$).

\section{Magnetic susceptibility of $S-$spin dimers}

The magnetism of two interacting $S-$spins, or dimers, has been
described quantitatively in the literature by the HDVV Hamiltonian
\cite{livro_khan}:

\begin{equation}\label{eq.01}
\mathcal{H} = -J\,\mathbf{S}_{A}\cdot \mathbf{S}_{B} -
g\,\mu_{B}\,\mathbf{B}\cdot\left(\mathbf{S}_{A}+\mathbf{S}_{B}\right)
\end{equation}
where $J$ is the exchange interaction, $\mathbf{S}_{A}$ and
$\mathbf{S}_{B}$ are the spins of each ion of the dimeric unit,
$g$ is the Land\'{e} factor, $\mu_{B}$ is the Bohr magneton and
$\mathbf{B}$ is an external magnetic field. When the system is in
thermal equilibrium, its thermal state is described by the density
operator $\rho = e^{-\beta\,\mathcal{H}}/Z$, in which $Z =
\mbox{tr}(e^{-\beta\,\mathcal{H}})$ is the partition function,
$\beta = 1/k_{B}\,T$ and $k_{B}$ is the Boltzmann constant. With
this operator, one can calculate, among other thermodynamical
quantities, the magnetic susceptibility, in which, for $S_{A}=
S_{B}$ and $\mathbf{B} \rightarrow 0$, holds as \cite{livro_khan}:

\begin{equation}\label{eq.02}
\chi(T) =
\frac{2\,\mathcal{N}\,(g\,\mu_{B})^{2}}{k_{B}\,T}\,\mathcal{F}^{(S)}(J,T)
\end{equation}
where, for $S=5/2$
\begin{equation}\label{eq.03}
\mathcal{F}^{(5/2)}(J,T) =
\frac{e^{x}+5\,e^{3\,x}+14\,e^{6\,x}+30\,e^{10\,x}+55\,e^{15\,x}}
{1+3\,e^{x}+5\,e^{3\,x}+7\,e^{6\,x}+9\,e^{10\,x}+11\,e^{15\,x}},
\end{equation}
$x = J/k_{B}\,T$ and $\mathcal{N}$ is the number of dimers. The
expression above describes the susceptibility of spin dimers with
$S_{A}=S_{B}=5/2$. For any $S_A =S_B$ value, ranging from 1/2 up
to 5/2, minor changes must be done in the equation above. For
$S_{A}=S_{B}=5/2-1/2=2$, one should suppress the last term of the
numerator and of the denominator. For $S_{A}=S_{B}=2-1/2=3/2$, one
should suppress again the other last terms, and so on, until
$S_{A}=S_{B}=1/2$, that is:

\begin{equation}\label{eq.04}
\mathcal{F}^{(1/2)}(J,T) = \frac{1}{3+e^{-x}}
\end{equation}

\section{Magnetic susceptibility as an Entanglement Witness}

Recent works have proposed the magnetic susceptibility as a
thermodynamical EW \cite{2007_PRB_75_054422, 2008_PRB_77_104402,
2009_PRB_79_054408, 2005_NJP_7_258, 2009_PRA_79_052337}. For a
system in which the Hamiltonian commutes with the $z$ spin
component, $[\mathcal{H},S_{z}]=0$, the average magnetic
susceptibility of $N$ $S-$spins in a complete separable state,
$\overline{\chi}(T)$, measured along the three orthogonal axis,
satisfies the relation \cite{2005_NJP_7_258, 2004_PRA_69_052327}:

\begin{equation}\label{eq.05}
\overline{\chi}(T) = \frac{\chi_{x}+\chi_{y}+\chi_{z}}{3} \geq
\frac{(g\,\mu_{B})^{2}\,N\,S}{3\,k_{B}\,T}
\end{equation}
The $EW$ is given by \cite{2005_NJP_7_258}:

\begin{equation}\label{eq.06}
EW(N) =
\frac{3\,k_{B}\,T\,\overline{\chi}(T)}{(g\,\mu_{B})^{2}\,N\,S}-1.
\end{equation}
Systems presenting $EW(N)<0$ are in an entangled state. Such an EW
has been used in recent works to detect entanglement in molecular
magnets \cite{2006_PRB_73_134404, 2006_PRA_73_012110,
2007_PRB_75_054422, 2008_PRB_77_104402}. In
Ref.\cite{2006_PRA_73_012110}, the magnetic susceptibility is
compared to a correlation function (see figure 1 of Ref.
\cite{2006_PRA_73_012110}) measured by neutron diffraction. The
correlation function shows entanglement in the same range of
temperature as the magnetic susceptibility, supporting the use
of the magnetic susceptibility as an entanglement witness.

From Eq.(\ref{eq.06}), together with Eq.(\ref{eq.02}), one can
identify the maximum temperature below which there is entanglement
between the spins of the dimers, for different spin values. This
temperature can be obtained from the inequality shown on Eq.(\ref{eq.07}), 
i.e., dimers with $S-$spin are in an entangled state if the inequality below is
satisfied:

\begin{equation}\label{eq.07}
\mathcal{F}^{(S)}(J,T)<\frac{S}{3}
\end{equation}
Considering that $x_{e} = J/k_{B}\,T_{e}$ and Eq.(\ref{eq.03}),
one can obtain the temperature of entanglement. For example, for
$S = 1/2$:
\begin{equation}\label{eq.08}
\mathcal{F}^{(1/2)}(J,T) < \frac{1}{6}
\end{equation}
and then
\begin{equation}\label{eq.09}
 T_{e}^{(1/2)}= -0.91\,\frac{J}{k_{B}}.
\end{equation}

\begin{figure}[t]
\unitlength1cm
\begin{minipage}[t]{7.5cm}
\begin{picture}(7.5,5.5)
\includegraphics[width=7.5cm,height=5.5cm]{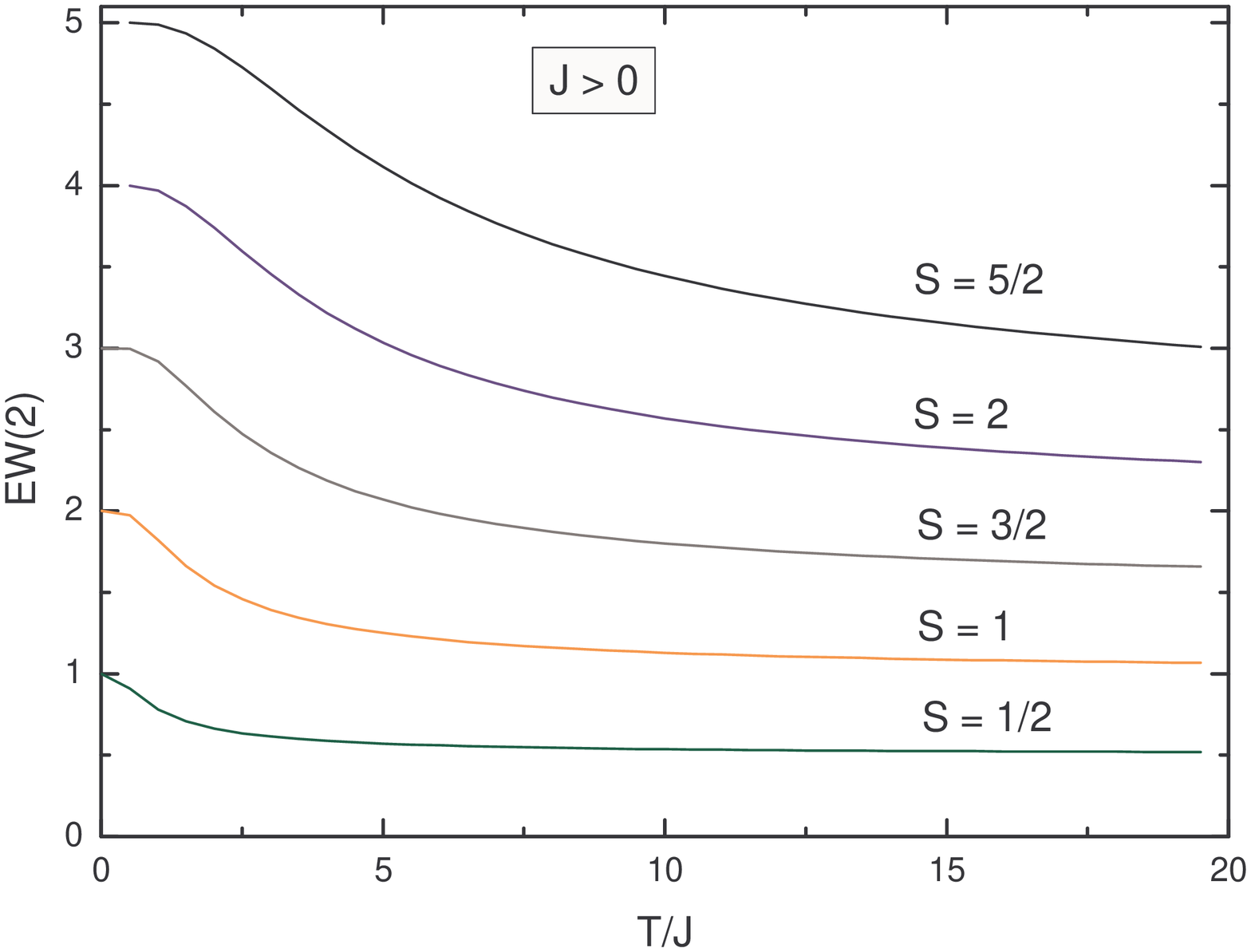}
\end{picture}
\end{minipage}
\vspace{0.1cm}
\begin{minipage}[t]{7.5cm}
\begin{picture}(7.5,5.5)
\includegraphics[width=7.5cm,height=5.5cm]{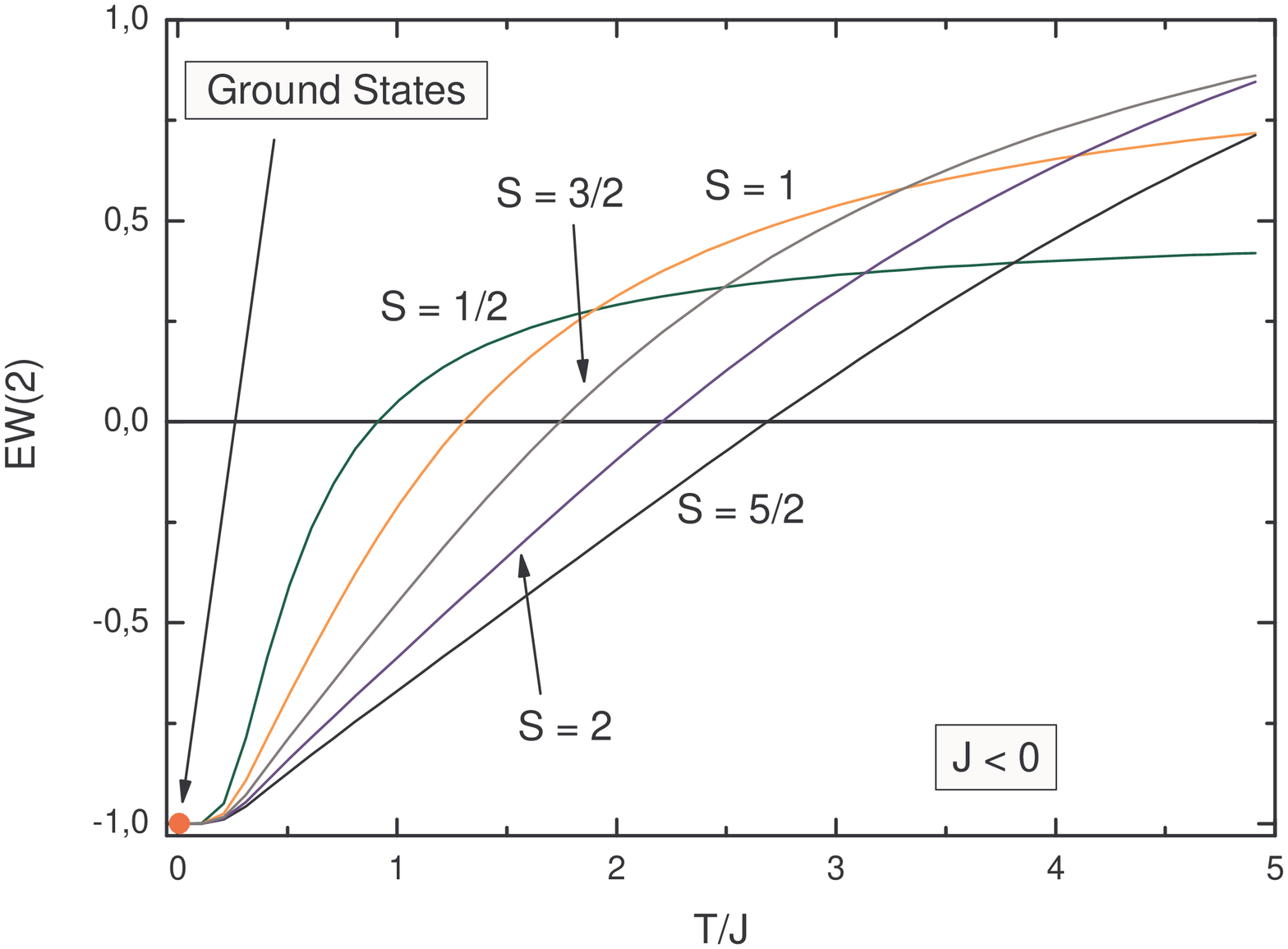}
\end{picture}
\end{minipage}
\caption{Numerical calculation of the $EW$, Eq.(\ref{eq.06}), for
ferromagnetic (upper figure) and antiferromagnetic (lower figure)
dimers. For the antiferromagnetic dimers, the temperature of
entanglement appears when $EW=0$. The entanglement in these
systems exists because its ground states are entangled. The red solid symbol
shows that $EW = -1$ for all the ground states in Table
\ref{table1}. On the other hand, the ferromagnetic dimers never
goes below zero.} \label{Fig1}
\end{figure}

Following this method, it is straightforward to obtain the
temperature of entanglement for the different spins. The temperatures of 
entanglement and the respective ground states are summarized in Table 1 for 
spins ranging from 1/2 to 5/2. In the limit $T
\rightarrow 0$, the Entanglement Witeness tends to $-1$, as shown in Fig.~\ref{Fig1}.
This happens because the ground states, shown in Table \ref{table1}, 
violate maximally the entanglement condition in Eq.(\ref{eq.06}). 
A similar calculation for entanglement temperature can be found in
Ref.\cite{2009_PRA_79_042334}, but for other spin$-1/2$ systems.

\begin{table*}
  \begin{center}
  \renewcommand\arraystretch{1.0}
  \begin{tabular}{|c|c|}
    \hline
      Temperature of entanglement & Ground state ($|m_{S_{A}},m_{S_{B}}\rangle$) \\\hline
     $T_{e}^{(1/2)} = -0.91\,\frac{J}{k_{B}}$ & $\frac{1}{\sqrt{2}}\left(|\frac{1}{2},-\frac{1}{2}\rangle - |-\frac{1}{2},\frac{1}{2}\rangle\right)$ \\
     $T_{e}^{(1)} = -1.30\,\frac{J}{k_{B}}$ & $\frac{1}{\sqrt{3}}\left(|1,-1\rangle - |0,0\rangle + |-1,1\rangle\right)$ \\
     $T_{e}^{(3/2)} = -1.74\,\frac{J}{k_{B}}$ & $\frac{1}{2}\left(|\frac{3}{2},-\frac{3}{2}\rangle - |\frac{1}{2},-\frac{1}{2}\rangle + |-\frac{1}{2},\frac{1}{2}\rangle - |-\frac{3}{2},\frac{3}{2}\rangle\right)$ \\
     $T_{e}^{(2)} = -2.21\,\frac{J}{k_{B}}$ & $\frac{1}{\sqrt{5}}\left(-|2,-2\rangle + |1,-1\rangle - |0,0\rangle + |-1,1\rangle - |-2,2\rangle\right)$ \\
     $T_{e}^{(5/2)} = -2.69\,\frac{J}{k_{B}}$ & $\frac{1}{\sqrt{6}}\left(-|\frac{5}{2},-\frac{5}{2}\rangle + |\frac{3}{2},-\frac{3}{2}\rangle - |\frac{1}{2},-\frac{1}{2}\rangle + |-\frac{1}{2},\frac{1}{2}\rangle - |-\frac{3}{2},\frac{3}{2}\rangle + |-\frac{5}{2},\frac{5}{2}\rangle\right)$ \\
    \hline
  \end{tabular}
  \end{center}
  \caption{Theoretical determination of the temperatures of entanglement $T_{e}$ for $S-$spin dimers.
  The ground states of the antiferromagnetic interacting dimers,
  written on the basis $|m_{S_{A}},m_{S_{B}}\rangle$, are entangled states
  as shown by the red solid symbol in Fig.~\ref{Fig1}. These ground states were numerically calculated and the eigenvalues, 
  with their respective eigenvectors, were determined. The case $S=2$ is included here for
  completness. In the actual experiment the coupling is ferromagnetic for this spin value (see text).}\label{table1}
\end{table*}

This result shows that the temperature below which the
entanglement emerges, or the temperature of entanglement, is
directly connected to the exchange interaction between the
$S-$spins of the dimer. Furthermore, as the exchange interaction
$J$ is strictly connected to the structure of the material, one
can affirm that entanglement can be ``engineered'' by adjusting
this quantity. In Fig.~\ref{Fig1}, one can see numerical
calculations for the Entanglement Witnesses of different spin
values. The results show that only the entangled state
of antiferromagnetic interacting dimers can be detected. This can
be understood considering the ground state of the dimers. For
antiferromagnetic interacting dimers, the ground state is an entangled 
state, thus maximally violating the criterion
presented in Eq.(\ref{eq.06}). On the other hand, for 
ferromagnetic interacting dimers, the ground state corresponds to a mixture of
entangled and non-entangled states and its respective EW do not indicate (or detect) entanglement. 
However, other entanglement witnesses are
capable of detecting entanglement in ferromagnetic states, as
discussed in Ref.\cite{2009_PRA_79_042334}. For that, other types
of experiments are needed.

\section{Experimental verification of the temperatures of entanglement}

In order to experimentally verify our theoretical prediction for
the temperature of entanglement, we have used the data obtained from the
magnetic susceptibility measurements of three isostructural
transition metal silicates with formula KNaMSi$_{4}$O$_{10}$ (M
$=$ Mn, Fe or Cu) \cite{2009_JSolStatChem_182_253}. For Copper
($S=1/2$) and Manganese ($S=5/2$) the interaction within the
dimers is antiferromagnetic and for Iron ($S=2$) a ferromagnetic
interaction holds \cite{2009_JSolStatChem_182_253}. These
materials cover our theoretical results, since they are dimers
with different $S$ values, different $J$ amplitudes and signals.

\begin{figure}[ht]
\begin{center}
\includegraphics[width=7.5cm]{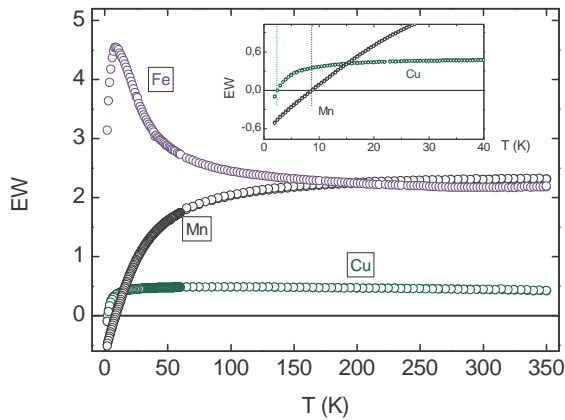}
\caption{Entanglement Witnesses for the three transition metal
oxides. For the Manganese and the Copper compounds
(antiferromagnetic interacting systems), the EW goes below zero,
showing the existence of entanglement among the spins. For the
Iron compound, the EW never goes below zero. Inset: Detail around
$T_{e}$. The orders of magnitude correspond to the ones
theoretically obtained (see Table \ref{table2}).} \label{Fig2}
\end{center}
\end{figure}

In Fig.~\ref{Fig2} it is shown the experimental determination
of the EW for the three compounds. As predicted in Table
\ref{table2}, only for antiferromagnetic interacting systems the
entanglement appears and the estimative of the temperature of
entanglement $T_{e}$ agrees with those obtained experimentally.

\begin{table*}
  \begin{center}
  \renewcommand\arraystretch{1.0}
  \begin{tabular}{|l|c|c|c|c|c|c|}
    \hline
                 & $J$ (K)  & $T_{e}^{theo}$ (K) & $T_{e}^{exp}$ (K) & Magnetic order\\\hline
    Cu ($S=1/2$) & $-2.86$ &   2.60   & 2.43(7) & antiferromagnetic \\
    Fe ($S=2$)   & 7.6     &  $-16.8$ & $-$     & ferromagnetic \\
    Mn ($S=5/2$) & $-3.83$ &  10.30   & 8.91(9) & antiferromagnetic \\\hline
  \end{tabular}
  \end{center}
  \caption{Comparison between the theoretical and experimental temperatures of entanglement $T_{e}$.
  It can be seen that one can only find a physical temperature if the interaction among the dimers is antiferromagnetic.
  Although the results are not exactly the same, one can have an estimative of the order of magnitude
  of the temperature of entanglement $T_{e}$ for different $S-$spin dimers.}\label{table2}
\end{table*}

\section{Conclusions}

In summary, we have shown a method to estimate the temperature of
entanglement in low dimensional magnetic materials composed of
$S-$spin dimers, for different spin values. We have found that 
$T_e \propto J$ and, therefore, the stronger the exchange 
interaction the higher the entanglement temperature. 
The results were experimentally verified using the magnetic susceptibility data from
Ref.\cite{2009_JSolStatChem_182_253} obtained for three
isostructural materials, KNaMSi$_{4}$O$_{10}$ (M$=$Mn, Fe
or Cu). The experiments reported in
Ref.\cite{2009_JSolStatChem_182_253} intended only to characterize
the basic magnetic properties of these materials. From these data
and the entanglement witness given in Eq.(\ref{eq.06}), the
existence of entangled states was determined, which proves 
that the theoretical model is good agreement with the experimental results. 
These results also allow affirming that this
EW can only detect entanglement in antiferromagnetic interacting
dimers. This happens because the ground state of ferromagnetic
interacting dimer is a mixture of entangled and non-entangled
states and this EW cannot distinguish them. On the other hand,
the ground state of an antiferromagnetic interacting dimer is an
entangled state. The theoretical prediction was experimentally
verified using three isostructural materials, which can be
theoretically modeled by a HDVV Hamiltonian. As one
possible direction for further studies, we suggest the investigation
of entanglement upon the increasing of the spin values. It is
important to mention that systems containing relatively large
spins can accurately be treated as classical Heisenberg spin
systems \cite{1999_PRB_60_10122}. The study of classical and
quantum spin clusters state and dynamics \cite{2000_PhysA_286_541,
2000_PhysA_278_214, 2007_NRL_2_168} can be very interesting for
the architecture of novel materials for quantum information
processing. Besides,  these studies allow the investigation of 
the role of the temperature in a  crossover between the quantum 
and classical descriptions of magnetic systems.

\acknowledgments The authors acknowledge T. G. Rappoport and 
T. J. Bonagamba for their comments. We would like to thanks the 
Brazilian funding agencies CNPq, CAPES and the Brazilian Millennium 
Institute for Quantum Information. D.O.S.P. acknowledges the 
financial support from FAPESP. A.M.S. acknowledges the government 
of Ontario. M.S.R. thanks the financial support from PCI-CBPF program.

\end{document}